# Absolute measurement of laser ionization yield in atmospheric pressure range gases over 14 decades

D. Woodbury[1], R. M. Schwartz[1], E. Rockafellow[1], J. K. Wahlstrand[2], and H. M. Milchberg[1,*]

[1]Institute for Research in Electronics and Applied Physics, University of Maryland, College Park, MD 20742, USA
[2]Physical Measurement Laboratory, National Institute of Standards and Technology, Gaithersburg, MD 20899, USA

Strong-field ionization is central to intense laser-matter interactions. However, standard ionization measurements have been limited to extremely low density gas samples, ignoring potential high density effects. Here, we measure strong-field ionization in atmospheric pressure range air, $N_2$ and Ar over 14 decades of absolute yield, using mid-IR picosecond avalanche multiplication of single electrons. Our results are consistent with theoretical rates for isolated atoms and molecules and quantify the ubiquitous presence of ultra-low concentration gas contaminants that can significantly affect laser-gas interactions.

The unification of tunneling ionization and multiphoton ionization (MPI) of atoms in intense laser fields by Keldysh in 1965 [1] provided an analytic foundation for strong field laser physics [2-6], but measurements of the transition from MPI to tunneling had to await later advances in short pulse lasers [7-10]. This transition is characterized in atomic units by the dimensionless Keldysh parameter $\gamma = (2\chi_p)^{1/2} \omega/E_0$, where $\chi_p$ is the atom's ionization potential, $E_0$ is the peak laser field, and $\omega$ is the laser frequency. At moderate intensity $I$ ($\gamma \gg 1$, MPI regime), the yield $Y$ is proportional to $E_0^{2n}$ ($\propto I^n$), while at higher intensities or longer wavelengths ($\gamma < 1$), the transition to tunneling and barrier suppression ionization [7,9] is characterized by $Y \propto I^{p<n}$, where $n$ is the integer number of photons needed to exceed $\chi_p$. These early measurements were conducted in extremely low density gases (typically $\sim 10^8 - 10^{12}$ cm$^{-3}$) in order to prevent ionization products interacting with background gas or experiencing space charge effects in transit to high voltage detectors [7,9,11]. However, many applications of strong-field ionization, such as high harmonic generation [12] or high intensity pulse propagation [13], occur at atomic densities many orders of magnitude higher where density-dependent ionization may be important.

Recent theoretical work, for example, suggests many-body effects in high density gases leads to an additional ionization channel: excitation-induced dephasing (EID) [14-17]. If "standard" isolated atom multiphoton/tunneling ionization is viewed as the result of optical-field induced dephasing of bound state-continuum coherence, which spoils the adiabatic following of the electron population in the strong, highly detuned optical field, then at elevated densities it was proposed that additional dephasing from Coulomb interaction with electrons in nearby atoms enhances ionization beyond the isolated atom process. Calculations [14] predict that the additional yield scales nearly linearly with density and is proportional to $I^2$ (in strong contrast with $I^n$ scaling for MPI), and is nearly independent of target species and laser wavelength. EID calculations predict yields $\sim 10^{-9}$-$10^{-7}$ cm$^{-3}$ for 1 TW/cm$^2$, $\lambda$=1-10 µm, 100 fs pulses in a variety of atmospheric pressure range gases [14-17], while at higher intensities the isolated atom rate dominates. While prior ionization yield measurements at atmospheric pressure [18,19] have shown reasonable agreement with isolated atom rates, they were limited to yields above $\sim 10^{-5}$ for a $\lambda$=800 nm driver, precluding investigation of EID ionization.

The potential effect of EID ionization is significant, especially when its boost to plasma density would have a commensurately larger effect on the refractive index experienced by longer wavelength lasers. For example, under conditions where standard ionization is negligible, EID was invoked to explain a recent experiment observing self-channeling of a $\lambda$=10.2 µm, ~1 TW/cm$^2$ peak intensity $CO_2$ laser pulse over 20 Rayleigh ranges in air [20], a process requiring plasma generation to offset Kerr self-focusing.

In this paper, we use avalanche ionization driven by a picosecond, mid-IR *probe* laser pulse to measure absolute ionization yields over 14 decades ($10^{-16}$ to $10^{-2}$) from femtosecond near-IR and mid-IR *pump* pulse irradiation of atmospheric pressure range air, nitrogen and argon (0.5-3 bar). This represents an unprecedented dynamic range with a single setup, with a sensitivity achievable by no other method we are aware of. In avalanche ionization, free electrons (here initially generated by femtosecond pump pulses) gain sufficient energy through probe-driven collisions until they ionize neutral atoms/molecules, leading to an exponential growth factor $e^{\nu_i t}$ in the local number of electrons, where $\nu_i$ is the effective collisional ionization rate. Growth saturates due to depletion of neutral molecules (for example, at 5% full single ionization of air, electron density ~$10^{18}$ cm$^{-3}$ and $\nu_i$ is reduced by 5%). Avalanche ionization was driven by a positively chirped ~10 mJ, 50 ps λ=3.9 μm probe pulse focused to intensities ~1-1.5 TW/cm$^2$ at a $1/e^2$ intensity radius (waist) of $w_0 = 70 \mu$m. The peak probe intensity defines a breakdown volume inside of which the intensity exceeds a threshold value leading to detectable avalanches (see supplementary material [21]). Crucially, the mid-IR avalanche driver eliminates driver-supplied MPI electrons from the seed population [22,23].

Figure 1 shows the experimental setup. The femtosecond pump pulse, synchronized to the avalanche-driving probe, was either in the near-IR (λ=1024 nm, 274±10 fs) or in the mid-IR (λ=3.9 μm, 85±5 fs) and focused to peak intensities of 1-100 TW/cm$^2$, with intensity control provided by a waveplate and polarizer. All three pulses were derived from a 20 Hz chirped pulse amplification (OPCPA) system [24], with details of focal spot measurements, breakdown volume, and absolute intensity uncertainty (~ ± 10% ) given in the supplementary material [22].

For low yields up to ~$10^{-11}$, visible avalanche breakdowns are local to individual seed electrons, with radial migration of avalanche-liberated electrons limited to < ~10 μm by electron and ambipolar diffusion during the 50 ps probe pulse [23,25]. Thus breakdowns are isolated and were counted by imaging, with a 16-bit low-noise CMOS camera, the overlap of the pump pulse and probe breakdown volume (Fig 1 (a)) inside a sealed gas cell filled with air, nitrogen, or argon passed through a 0.01 micron rating particulate filter. In this regime, the occurrence of breakdowns is statistical, requiring multi-shot averaging. In order to determine peak yield $Y_0$ corresponding to the peak intensity $I_0$, we use $N = \int_V Y_0 \times (I(r,z)/I_0)^m \, dV$, where $N$ is the average number of counts measured, $I(r,z)$ is the spatially varying pump intensity with peak value $I_0$ over the probe breakdown volume $V$ , and the yield is observed to scale as $I^m$. A counter-propagating ($\theta = 0°$) pump-probe geometry maximized the overlap volume and hence sensitivity. As higher pump intensity increased the number of seed electrons beyond ~10, individual breakdowns upstream interfered with probe driving of downstream avalanches. Switching to a perpendicular geometry ($\theta = 90°$) reduced the overlap volume ~100 ×, eliminating this propagation effect at higher yield. While the small volume for $\theta = 90°$ prevents reliably imaging more than 1 breakdown per shot, counting the incidence of no breakdowns allowed us to infer the Poissonian mean up to ~4 breakdowns/shot, since a Poisson distribution with mean value $\mu$ has a probability $P(0) = 1 - e^{-\mu}$ of observing no counts. With the pump blocked, breakdowns occurred in ~1 out of 100-1000 shots due to probe-induced MPI of a contaminant (see below).

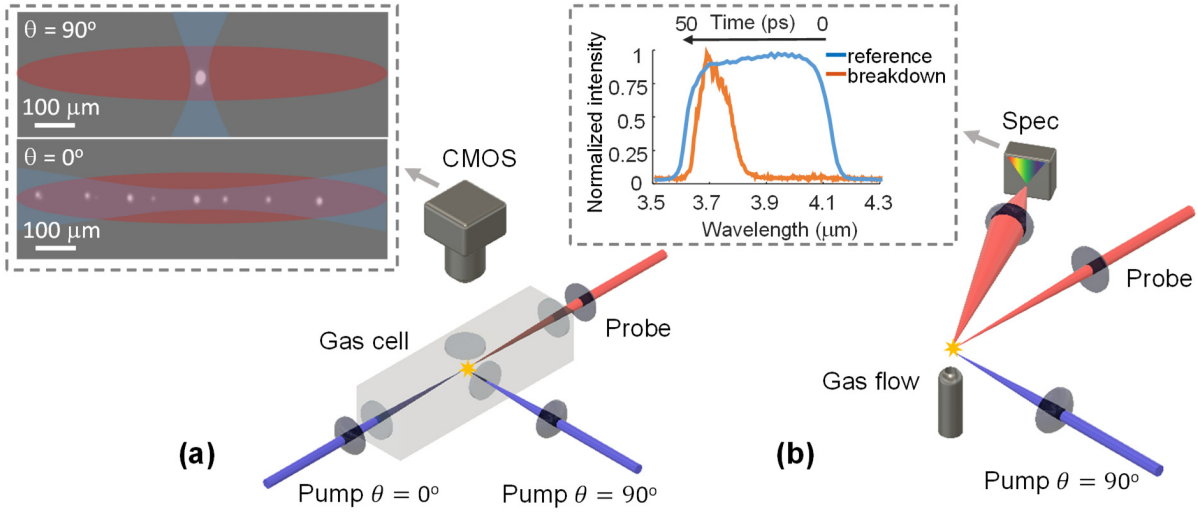

**FIG. 1 | Experimental setup.** (a) *Breakdown counting* ($I < 10$ TW/cm$^2$): A positively chirped, λ=3.9-4.2µm, 50 ps mid-IR laser probe pulse was focused into a gas cell to drive avalanche breakdowns seeded by electrons liberated by either a counter-propagating ($\theta = 0°$) or a perpendicularly-directed ($\theta = 90°$) pump pulse (274 fs, $\lambda = $ 1024 nm or 85 fs, $\lambda = 3.9 \ \mu$m). The inset shows, for each geometry, sample images of individually seeded breakdowns, collected by camera CMOS, and overlaid with pump pulse focal volume (blue) and the probe pulse breakdown threshold volume (red). (b) *Breakdown time advance* ($I > 10$ TW/cm$^2$): pump-induced initial plasma density and corresponding yield are determined from breakdown timing encoded in the backscatter spectrum of the chirped mid-IR probe pulse. Backscatter is collected by spectrometer Spec, with example incident and backscattered spectra and corresponding timing shown. Here, breakdowns are observed directly above a ~5 mm gas flow orifice.

As the yield (and seed electron density) increases even further, ≲$10^{-10}$ to $10^{-2}$, adjacent incipient avalanche sites become closer than the electron diffusion length and it is no longer possible to resolve and count breakdowns. However, the avalanche is now seeded by a well-defined local electron density such that one can measure a deterministic avalanche time, $\tau = \ln(N_{ef}/N_{e0})/\nu_i$, where $N_{e0}$ is the seed electron density, $\nu_i$ is the electron density collisional growth rate, and $N_{ef}$ is a final (detectable) electron density [26, 27]. By employing our previously demonstrated chirped probe-backscatter breakdown timing method [23], we measure the breakdown time advance $\Delta t_{adv} = \tau_{driver} - \tau$, where $\tau_{driver} = 50ps$ is the avalanche driver duration, and $\Delta t_{adv}$ corresponds to the reddest (earliest) wavelength of the chirped probe pulse detectable in the backscattered spectrum at a detection threshold $N_{ef} \sim 10^{18}$ cm$^{-3}$. The spectrum is collected by a single shot mid-IR spectrometer [23], with setup and example spectra shown in Fig.

1(b). Wavelength-to-time correspondence of the chirped driver was established using a cross correlation with the λ=1024 nm beam.

Figures 2–4 together show femtosecond pulse ionization yields $Y$ spanning 14 orders of magnitude. For lower peak intensities of $0.6 - 10$ TW/cm², where yields are determined from counting individual breakdowns, Fig. 2 plots $Y_{1024nm}$ for air (a), a comparison of $Y_{1024nm}$ for air, N$_2$, and Ar (b), and $Y_{3.9\mu m}$ for air (c), all at atmospheric pressure. Here, $\gamma_{1024nm} > 3$ and $\gamma_{3.9\mu m} < 0.9$, in the MPI and tunneling regime, respectively. The corresponding average breakdown counts/shot are shown on separate scales. In 2(a), the curves for $\theta = 0°$ and $\theta = 90°$ are horizontally offset owing to peak intensity uncertainty of $\sim \pm 10\%$ (horizontal bars) in each geometry [21]. Theoretical isolated molecule yields were calculated using a rate valid for arbitrary $\gamma$ by properly treating the Coulomb correction in the multiphoton limit $\gamma \gg 1$ [2,6]. This "standard" yield for air (80/20 N$_2$/O$_2$) and N$_2$, using effective potentials for N$_2$ and O$_2$ [10], is plotted as the yellow curves in Fig. 2(a, c), with the curve in Fig. 2(c) scaled up by 100×.

Best fits to the data points for all 3 gases give $Y_{1024nm} \propto I^{5.5 \pm 0.3}$ (for $I < \sim 4$ TW/cm²) and $Y_{3.9\mu m} \propto I^{12.1 \pm 0.8}$, with measured yields orders of magnitude greater than standard theory. For air at $I > \sim 4$ TW/cm², the yield dependence transitions to $Y_{1024nm} \propto I^{9.7 \pm 1.0}$, consistent with the expected MPI scaling of $I^{10}$ for oxygen, the most readily ionized air constituent ($\chi_p \sim 12.1$ eV). The $\sim 2 \times$ offset between experiment and theory in this range is consistent with $\sim 10\%$ experimental uncertainty in absolute intensity [21] and the lack of species-specific atomic structure in the theoretical rate [6]. The range of exponents is determined by the 95% confidence interval for linear fitting to data on a log-log scale.

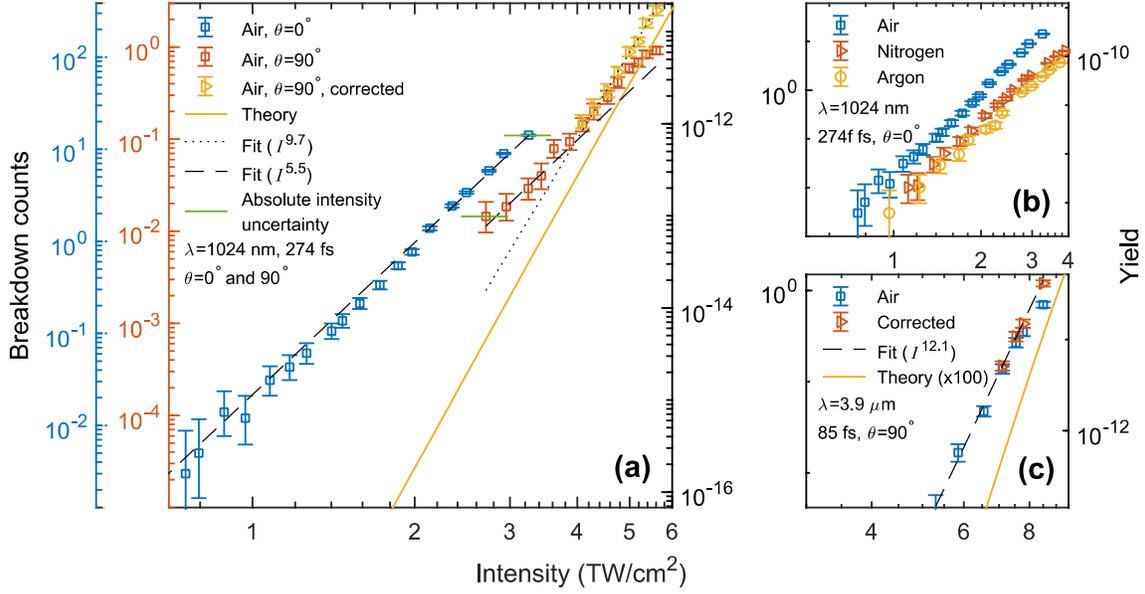

**FIG. 2** Ionization yield measured in breakdown counting regime ($I <10$ TW/cm$^2$). **(a)** Breakdown counts and corresponding yields $Y_{1024nm}$ in $\theta = 0°$ and $\theta = 90°$ geometry. For $I > \sim 4$ TW/cm$^2$, $Y_{1024nm} \propto I^{9.7}$, consistent with MPI of oxygen ($\chi_p \sim 12.1$ eV) and for $I < \sim 4$ TW/cm$^2$, $Y_{1024nm} \propto I^{5.5}$, consistent with MPI of a contaminant with $\chi_p \sim 6$ eV. Error bars correspond to a Poissonian 95% confidence interval [28]. Horizontal bars on the 0° and 90° plots reflect absolute intensity uncertainty from switching between geometries. The overlaid theory curve plots the yield based on standard N$_2$ and O$_2$ molecular ionization rates [6,10]. **(b)** Comparison of $Y_{1024nm}$ for atmospheric pressure air, N$_2$ and Ar for $I < \sim 4$ TW/cm$^2$ ($\theta = 0°$), showing $Y_{1024nm} \propto I^{5.5}$ for all 3 gases. **(c)** $Y_{3.9\mu m}$ for atmospheric pressure air ($\theta = 90°$). The overlaid ionization theory curve for 80/20 N$_2$/O$_2$ is multiplied by 100. In (a) and (c), saturated counts were corrected statistically, as described in the text.

These results strongly suggest that the ionization yield at lower intensity originates from a contaminant common to all three gases. Further supporting the presence of a contaminant, when the cell was filled with bottled, high purity air passed through a Supelcarb part-per-billion level hydrocarbon trap, the ionization yield dropped by a factor of ~4, with the intensity scaling remaining the same [21]. The gas cell experiment was repeated for air pressures of 0.5 bar to 3 bar. At all pressures, the yield scaling at lower intensity followed $Y_{1024nm} \propto I^{5.5 \pm 0.3}$, consistent with the presence of the contaminant. At higher intensity, the yield dependence transitioned to the MPI scaling of oxygen, as in Fig. 2(a).

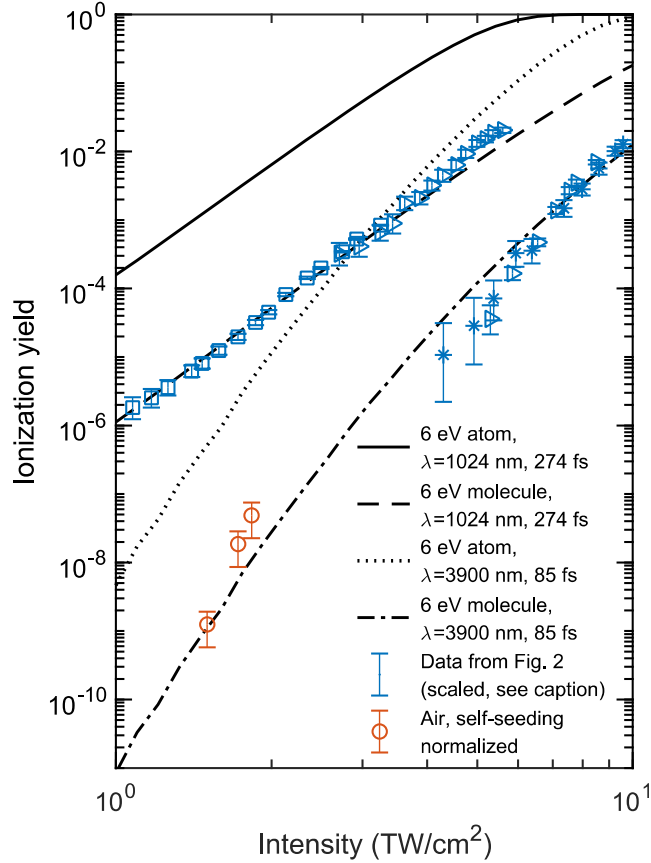

**FIG. 3** Comparison of contaminant yield with isolated atom/molecule theory. Counts in air ($\theta = 0°$, squares; $\theta = 90°$, triangles) and argon ($\theta = 90°$, stars) are shifted to overlap with theoretical curves for comparison. The red points (open circles) are from 50 ps probe self-seeded avalanches, normalized to the other λ=3.9 μm data on the plot (see Fig. S2 and discussion in [21]). Two theory curves were generated for each wavelength: a 6 eV atom and a 6 eV molecule. The molecule calculation employs a shape factor (0.5) to account for changes to its tunneling potential [10]. The calculated contaminant yields, together with the absolute yields determined in Fig. 2, suggest a contaminant concentration in the range $10^{-11}$-$10^{-9}$, assuming a shape factor bounded between 1 (atom) and 0.5 (typical molecule).

As shown in Fig. 3, fitting the yield scaling to the standard isolated atom/molecule ionization rate [6] suggests the contaminant species has an ionization potential $\chi_p$~6 eV and an approximate concentration of ~$10^{-11}$–$10^{-9}$. The ≲2 TW/cm² (red) points with $Y_{3.9\mu m} \propto I^{19\pm8}$ were obtained from counting breakdowns self-seeded by the λ=3.9 μm probe and were normalized to short pulse results as described in [21], and are also consistent with MPI of a $\chi_p$~6 eV contaminant. We note that early MPI experiments indicated the presence of low ionization potential contaminants in all laboratory gases; these were considered to be the source for seed electrons in air avalanche breakdown experiments [7,11]. However, the concentration and yield of these seed sources could not be quantified as in the present work.

Figure 4 covers the transition from MPI of air and $N_2$ to tunneling ionization, with $3 > \gamma_{1024nm} > 0.75$. In this regime, we used our breakdown time advance diagnostic. Conversion from $\Delta t_{adv}$ to yield was calibrated by data from the direct imaging measurements at ~6 TW/cm$^2$ (Fig. 3(a)) and previous absolute measurements of yield at ~100 TW/cm$^2$ [18], with direct interpolation between the points assuming a constant growth rate for a flat-top probe pulse intensity, as explained in greater detail in [21]. Measured yields and theory show agreement within a factor of 10 over the full intensity range despite the simplistic assumption of constant growth rate. Accounting for the probe pulse temporal envelope and chirp-dependent heating would bring the curves into even closer agreement [21]. We note that the growth rate, $v_i = 0.55 \text{ ps}^{-1}$, extracted from this interpolation also applies to laser-air interactions with a different wavelength but the same numerical value of $I\lambda^2$, and can be used to benchmark simulations of high intensity, picosecond laser-driven avalanche.

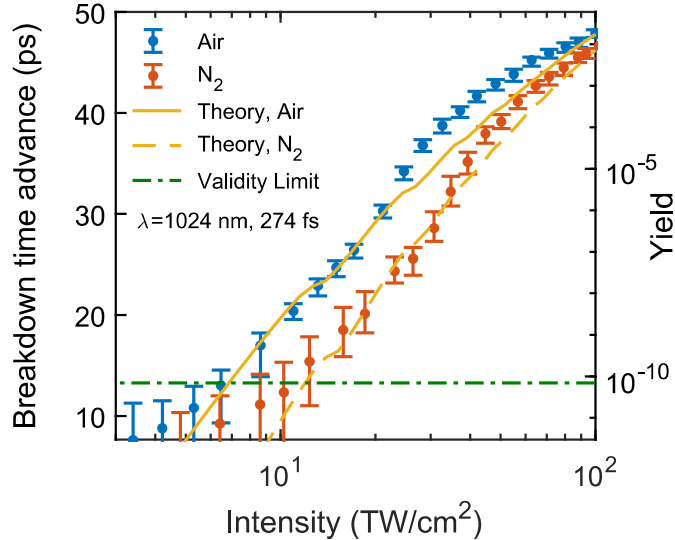

**FIG. 4** *Ionization yield measured in breakdown time advance regime ($I$ >10 TW/cm$^2$).* Ionization yields in air and $N_2$ determined by breakdown time advance $\Delta t_{adv}$, with theoretical yields overlaid. The horizontal dot-dashed line indicates the limit above which $\Delta t_{adv}$ is deterministically correlated with seed electron density. Below this level, individual breakdowns do not overlap during their initial growth phase, and breakdown timing is determined by statistical placement of seed electrons in the pump volume (Fig. S4 in [21]). Conversion to yield was benchmarked with imaging results from Fig 2(a) and previous measurements [18]. The points at each intensity give the mean $\Delta t_{adv}$, and error bars show the standard deviation of timing measurements due to either statistical placement of seeds (low intensity) or ~5% fluctuations in probe pulse energy (higher intensity).

Our femtosecond pump ionization yield measurements can be summarized as follows: At lower intensities (Fig. 2 and Fig. 3, $< \sim 4 \text{ TW/cm}^2$) where the biggest relative ionization contributions from EID are expected, pump wavelength-dependent scaling of yield is consistent with ionization of a low-level contaminant with $\chi_p \sim$ 6 eV. The $I^2$ scaling suggested by EID is not observed, even when measuring yields $10^6$ lower than those predicted [15-17]. The disagreement with the EID theory may arise from its assumption of electron delocalization in the atmospheric density regime and potential self-action (see supplementary discussion in [21]). In the higher intensity range $4 \text{ TW/cm}^2 < I < 10 \text{ TW/cm}^2$ of Fig. 2, the yield at λ=1024 nm transitions into MPI of $O_2$, while it is in the tunneling regime of the contaminant for λ=3.9μm. In Fig. 4, at higher intensities up to 100 TW/cm$^2$ and the transition from MPI to tunneling, the ionization yield is in good agreement with isolated atom/molecule theory.

We note that avalanches seeded by low ionization potential contaminants could have a significant effect on long wavelength infrared filamentation and be consistent with the observations of self-channeling of $\lambda = 10.6$ μm, TW-level $CO_2$ laser pulses [20] without the need for EID ionization. Not only can a long-wave IR pulse easily ionize the $\chi_p \sim$ 6 eV contaminant, but the $\lambda^2$ dependence of collisional heating and free electron polarizability [17,20,29] renders such a pulse quite sensitive to any free electrons it self-generates and their subsequent avalanche growth.

In conclusion, we have shown that avalanche breakdown using picosecond mid-IR probe pulses is a sensitive diagnostic of extremely low electron densities—achieving an unprecedented dynamic range of 14 orders of magnitude, with picosecond and few micron resolution. We measure ionization generated by femtosecond pump pulses in several gases in the atmospheric pressure range and find that the yield at lower laser intensities $\sim 1 \text{ TW/cm}^2$ is consistent with MPI of a ubiquitous parts-per-trillion contaminant, and is not dependent on predicted many body effects, while yield at higher intensities ($> \sim 10 \text{ TW/cm}^2$) the yield is consistent with MPI or tunneling ionization of isolated molecules. In particular, our avalanche method enables measurement of intermediate electron densities in a range ($10^8$-$10^{13}$ cm$^{-3}$) inaccessible with other standard techniques without sacrificing spatial or temporal resolution [30-34].

**Acknowledgements**

The authors thank J. Isaacs and P. Sprangle for useful discussions and for assistance in simulating avalanche breakdowns.

This work is supported by the Air Force Office of Scientific Research (AFOSR) (FA9550-16-1-0121, FA9550-16-10259), Office of Naval Research (ONR) (N00014-17-1-2705), Defense Threat Reduction Agency (DTRA) (HDTRA11510002), and the National Science Foundation (NSF) (PHY1619582). D. W. acknowledges support from the DOE NNSA SSGF program under DE-NA0003864.

*Corresponding author: milch@umd.edu

**Supplementary Note 1: Beam measurement and avalanche breakdown volume**

The λ=1024 nm pump pulses were focused to Gaussian waists ($1/e^2$ intensity radius) of $w_0$=8 μm ($I < 100$ TW/cm², Rayleigh range $z_0$~0.2 mm) or $w_0$= 26 μm ($I < 10$ TW/cm², $z_0$~2 mm) for $\theta = 90°$, while for $\theta = 0°$, they were focused to $w_0$= 30 μm. The λ=3.9μm pump pulses were focused to $w_0$=39 μm ($z_0$~1.2 mm, up to 10 TW/cm²) for $\theta = 90°$. Near-IR and mid-IR peak pump intensities were determined by measuring focal spots directly on a CCD camera or an InSb array, respectively. Pulse duration measurements made with using an autocorrelator (for λ=1024 nm) or with scanning second-harmonic generation frequency resolved optical gating, or SHG-FROG (for λ=3.9 μm). Uncertainty in pulse duration ($\pm\sim 5\%$) and focused beam spot size ($\pm\sim 4\%$ due to finite pixel size) gives absolute uncertainty of $\sim \pm 10\%$ in measured intensity values. Breakdown occurs if electron growth from heating and subsequent ionization exceeds losses due to recombination, attachment, and diffusion out of the laser focal volume, leading to a characteristic intensity threshold [1-4]. For short pulses, this breakdown criterion is increased in order to drive avalanche to some detectable threshold before the end of the pulse, which in this case was detection of a visible breakdown site in images of the interaction region. Images were collected at 2× magnification on the CMOS camera, and the number of breakdowns was determined by counting the number of sites with peak signals above 20 pixel counts after median filtering. In order to determine the breakdown threshold for different gases and pressures, probe pulse peak power was reduced until the pump-seeded breakdowns at the center of the probe volume (peak probe intensity) were barely visible (~20 pixel counts). This gave a breakdown threshold $I_{th}$~1 TW/cm² in nitrogen and air, and 0.6 TW/cm² in argon, with a $1/p$ pressure dependence for all gases studied, in line with past observations [1]. When counts were converted to yield during data collection (peak intensity typically ~$1.5 I_{th}$), integration was performed over the volume of the probe beam where $I > I_{th}$, as determined by direct measurements of the probe beam waist and the beam longitudinal profile.

**Supplementary Note 2: Effect of filter on contaminant breakdowns and scaling of self-seeded (probe produced) breakdowns**

When air passed through the particulate filter was replaced with bottled ultra-high purity air passed through an activated charcoal Supelcarb hydrocarbon filter (capable of filtering primary hydrocarbons to ~the part-per-billion level) for identical pump conditions ($\lambda = 3.9\ \mu m, \theta = 90°$), the number of breakdown counts decreased by ~4×, as shown in Fig. S1.

Since counts could not be observed below ~5 TW/cm² in this configuration and pumping in the $\theta = 0°$ geometry with the 3.9 $\mu m$ pump was experimentally difficult, low yield ionization was tested using breakdown counts initiated by "self-seeded" electrons, or seed electrons produced by the leading edge of the 50 ps, ~1.5 TW/cm² probe pulse which were subsequently amplified and detected as breakdown counts generated by the remainder of the pulse. Figure S2 shows both raw counts for 1.5-1.8 TW/cm², and corrections for the increase in size of the breakdown volume with increasing intensity and the increase in the time during which electrons can be liberated and amplified. Namely, if single electron-seeded breakdowns at a local intensity of 1.8 TW/cm² occur in 27 ps while breakdowns at 1.5 TW/cm² occur in ~33 ps, the second pulse has 6 ps longer in which to ionize contaminants through MPI and still drive a detectable breakdown. While the correction is simplistic (applying changes in volume and timing as a constant multiplicative factor and ignoring spatial variations in yield and timing), it gives a rough estimate of the scaling in this regime. In particular, it shows that counts are still driven by MPI/tunneling, with a best fit of the corrected counts giving $Y_{3.9\mu m} \propto I^{19\pm 8}$. To incorporate this data into Fig. 3 of the main text, the data was normalized to the data taken with the 85 fs pump pulse by accounting for the ratio in volumes between the self-seeding case and the $\theta = 90°$ geometry (~300) and the change in pulse length and temporal shape (~1000×). This normalization gives a yield in reasonable agreement with the theoretical curve, and consistent with a contaminant with ionization potential ~6 eV.

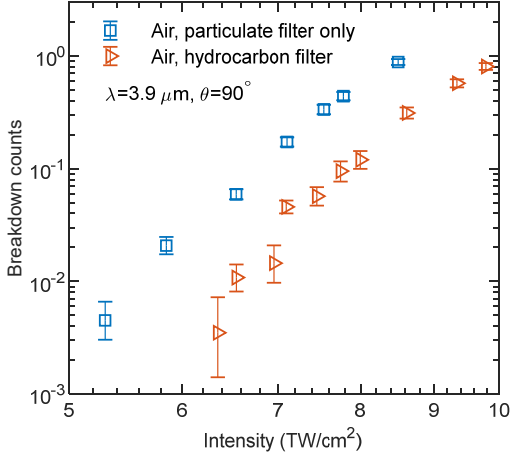

**Fig. S1 | Effect of hydrocarbon filter on yield measurements in air.** Breakdown counts observed with the λ=3.9 μm pump for two sources of air: (i) air passed through only the particulate filter or (ii) ultra-high purity compressed air from a bottle fitted with an additional part-per-billion hydrocarbon trap.

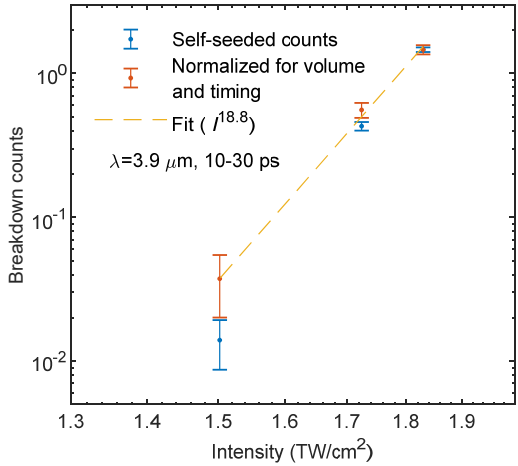

**Fig. S2 | Scan of self-seeded counts.** Number of breakdowns observed for varying probe (only) pulse intensities, with seed electrons liberated via MPI by the leading edge of the pulse. Also shown is a correction for changes in breakdown volume and effective seed timing as probe intensity is increased. A best fit of the points gives scaling $Y_{3.9\mu m} \propto I^{19\pm 8}$, with the large uncertainty set by the limited range of results.

**Supplementary Note 3: Avalanche breakdown timing, simulations, and density estimation from time advance measurements**

For a single electron seed, the time required to reach a specific breakdown condition is determined entirely by the local intensity, which in turn determines electron heating, temperature, and growth rate. However, if only a few seed electrons are randomly placed within the breakdown region, the time required to reach breakdown will show statistical variation due to spatial variations in intensity [5]. As the density of seeds is increased, there is a high probability of an electron being found in the region of highest intensity, leading to a deterministic breakdown time. As mentioned in the main text, once two seed electrons are closer than the diffusion length, the time required to reach saturation will decrease further, since the number of doublings in electron number (generations) will be reduced [2,6].

Breakdown timing was measured by observing the spectrum of pump light backscattered from the interaction region. Since the pump pulse is positively chirped to a length of 50 ps FWHM (70 ps full width) from its bandwidth-limited duration of ~80 fs, each spectral component corresponds to particular time slice in a 70 ps window. Energy backscattered from the plasma is detected with a single-shot mid-IR spectrometer, with a cryogenically cooled InSb detector, with a minimum electron density of ~$10^{17}$-$10^{18}$ cm$^{-3}$ required for detection of the backscatter signal, based on the analysis in [7] and its supplementary material. The longest wavelength above the detector noise threshold was recorded for each shot, and then breakdown timing was determined by the time-frequency mapping determined through a cross correlation measurement.

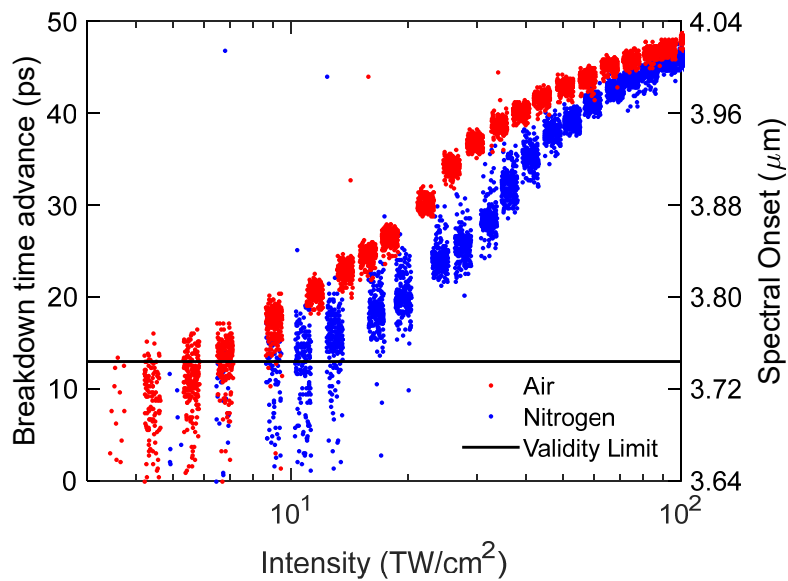

**Fig. S3 | Single shot breakdown timing.** Each point corresponds to a single probe pulse backscattered spectrum measurement, with the right vertical scale showing the longest wavelength detected, and the associated breakdown time advance shown on the left vertical scale. For low pump intensities, liberated electrons, when they are generated at all, are randomly positioned in the probe breakdown volume, leading to a spread of breakdown times. As pump intensity is increased, multiple seed electrons are generated and more are likely to be found at the peak probe intensity, which visually corresponds to ~13 ps time advance. As more breakdowns occur, they begin to overlap, leading to a deterministic decrease in breakdown timing (>13 ps advance), with the spread in points in that part of the plot determined by fluctuations in probe intensity.

For the present experiment, the width of the breakdown volume (region above threshold) for driving avalanche and backscattering was ~40μm, while the diffusion-limited diameter of single-electron-seeded breakdown plasmas was ~10μm during their initial growth phase. Thus, even when a single breakdown occurs on every shot, timing measured by backscatter will be variable because a single pump-generated seed electron could find itself in a range of intensities above the breakdown threshold. This is seen in Figure S3, where the points at ~<6 TW/cm$^2$ show a timing variation even though there is a ~1 breakdown/shot at that intensity. Once ~4-8 seed electrons are distributed in the breakdown region, there is a higher probability that one seed electron is located at the region of peak probe intensity, leading to more deterministic timing. Above ~8-10 seed electrons in the breakdown region, there is a high probability of 2 electrons being located at the region of peak intensity and within ~10μm of each other, such that number of generations needed to reach the detection threshold is reduced by one. This leads to our estimate that time advance is directly correlated with density for yields above $7 \times 10^{-11}$cm$^{-3}$, namely 10 times the yield corresponding to ~1 breakdown per shot calculated in Fig. 2 of the main text. On the plot of time advance, this density is then used to match the time advance where statistical breakdown ends, as shown in Fig. S3. Yields at 100 TW/cm$^2$ were matched exactly with the standard theoretical rate [8], since measurements of O$_2$ and N$_2$ yield in a thin gas jet with a 42 fs, 800 nm pulse at this intensity [9] showed excellent agreement with the theoretical rate for this same intensity range. Interpolating between the two gives an electron density growth rate of $\nu = 0.55$ ps$^{-1}$ during the probe pulse, which was used to calculate the intermediate densities. We note that our chirped probe pulse temporal profile is not exactly square, with a spectral measurement of the OPCPA's mid-IR beam and its near-IR conjugate suggesting more power at the beginning of the pulse, so that ponderomotive heating ($\propto I\lambda^2$) will be stronger at the beginning of the chirped pulse than at the end. Accounting for this would tend to suppress the inferred density slightly throughout the range, bringing it into closer agreement with the theoretical rate.

It is worth noting that the backscattering method does not rely on simulations, which in turn are dependent on accurate rates for elastic and inelastic collisions, attachment, diffusion, and transport. Nevertheless, a comparison with simulations can give confidence in the general approach. Using a constant intensity of 1.3 TW/cm$^2$ (the peak probe intensity used for high yield measurements) in a self-consistent set of 0-D equations that track the temperature of avalanching electrons through electron-neutral collisions (heating), and attachment, excitation, dissociative and ionization losses, [2,6] predicts a growth rate of $\nu = 0.35$ ps$^{-1}$ after ~2 ps of initial heating needed to reach a steady state plasma temperature of 10 eV. This matches reasonably well with the growth rate assumed by observing the 35 ps change in initial density of $2 \times 10^8$ from Fig. 4 of the main text, which gives a growth rate of $\nu = 0.55$ ps$^{-1}$. We note that the simulations are sensitive to uncertainty in loss rates and collision rates, as well as any departure from the assumption of a thermal electron distribution, so disagreement is not unexpected.

**Supplementary Note 4: Evaluation of theory of ionization through excitation-induced dephasing**

Excitation-induced dephasing (EID) describes dephasing in a quantum system induced by an excited state population. It is a well-established phenomenon in semiconductor materials [10]. In the theory of [11-13], EID from many-body effects is responsible for enhanced ionization because it disrupts adiabatic following, which causes the transient continuum population produced by a strong optical field to return to the ground state every half cycle. Adiabatic following relies on the induced polarization being phased properly with respect to the optical field. Dephasing of this polarization results in an accumulating residual population in the continuum.

In [11-13], the randomly positioned atoms in the gas (not moving on the optical pulse time scale) are treated as a homogenous density of atoms similar to "jellium" models. The bound electronic state of each atom is approximated by a single state, and the continuum states are treated in a reciprocal state basis. The resulting set of differential equations (Eqs. (4-7) in [11]) for the populations (i.e. the diagonal matrix elements of the density matrix) and polarizations (off-diagonal matrix elements) of the bound state and continuum states is very similar to the semiconductor Bloch equations [14].

In [13], it is argued that interaction between continuum electrons is the most important many-body effect, so it is what we focus on here. The many-body term responsible for dephasing (Eq. (12) in [11]) is a sum over continuum states indexed by wavevector **k**. Each basis state is delocalized. However, it is well established [15] that the electron wavepacket during strong field interaction is localized near the atom (particularly in the relatively low intensity range relevant here). Over an optical cycle time scale, there is simply not enough time for the electron to get very far from the atom. The gas is composed of hydrogen atoms, so any many-body interaction between electrons has to involve electrons from one atom interacting with electrons from different atoms. In treatments of electron-electron interactions in the solid state, it is necessary to exclude the interaction of an electron with itself (the "self-interaction") (see Chapter 7 in [14]). Excluding this in the hydrogen atom gas treated in [11-13] would necessarily involve excluding electron states corresponding to the same atom, and this information seems to have been lost in the approximation that all atoms can be treated as a homogeneous medium. The sums over wavevector difference **q** could exclude **q** = 0 as is done in solid state context (though this is not explicitly stated in [11-13]), but that would be insufficient for excluding self-interaction, because an electron wavepacket near a hydrogen atom is composed of many **k** states. It is possible that self-interaction was excluded in the derivation of the many-body term in some other way, but no exclusion of self-interaction is mentioned in [11-13]. Including self-interaction would likely lead to a dramatic overestimate of EID-induced ionization.